\documentclass[epj]{svjour}
\usepackage{graphics}
\newcommand{\ds}{\displaystyle}
\newcommand{\dsf}{\ds\frac}
\newcommand{\Tr}{\mbox{Tr}}
\newcommand{\re}[1]{(\ref{#1})}
\newcommand{\sF}{{S}}

\begin{document}
\title{Neutron-proton mass difference in  nuclear matter}

\author{Ulf-G. Mei{\ss}ner\inst{1,2} \and A.~M.~Rakhimov\inst{3}
\and A.~Wirzba\inst{2} \and U.~T.~Yakhshiev\inst{2,4}
}                     % Do not remove
\institute{
Helmholtz-Institut f{\" u}r Strahlen- und Kernphysik (Theorie),
D-53115, Universit{\" a}t Bonn, Germany
\and
Forschungszentrum
J{\" u}lich, Institut f{\" u}r Kernphysik
(Theorie),  D-52425  J{\" u}lich, Germany 
\and
Institute of Nuclear Physics, Academy of Sciences of
Uzbekistan, Tashkent-132, Uzbekistan
\and
Physics Department \& Institute of Applied Physics,
National University of Uzbekistan, Tashkent-174, Uzbekistan}
\date{Received: date / Revised version: date}
\abstract{
Isospin-breaking effects in nuclear matter are studied
in the framework of a medium-modified Skyrme model. The proposed
effective Lagrangian incorporates both the medium influence of
the surrounding nuclear environment on the single nucleon properties
and an explicit isospin-breaking effect in the mesonic
sector. The approach predicts that the
neutron-proton mass difference decreases in isospin-symmetric nuclear
matter but by a very small amount only.
 % end of PACS codes
} %end of abstract
\maketitle

\section{Introduction}
\label{sec:intro}

The evaluation of isospin-breaking effects in a dense nuclear medium is
an interesting problem in nuclear physics. This is 
particularly the case for the Nolen-Schiffer anoma\-ly~\cite{Nolen:1969ms} in
mirror nuclei  which may be explained  by a change of the
neutron-proton mass difference in the nuclear
environment. Henley and Krein~\cite{Henley:1989vi} argued that the
Nolen-Schiffer anomaly can be resolved if 
the neutron-proton mass difference decreases rapidly in the medium.
For this reason various approaches for
calculating the neutron-pro\-ton mass difference in a dense nuclear
environment have been proposed in the 
literature~\cite{Hatsuda:1990pj,Hatsuda:1990zj,Adami:1991js,Schafer:1993xt,Saito:1994tq,Meissner:1991if,Fiolhais:1991qf}.
Some of these approaches indeed confirmed a
decrease~\cite{Hatsuda:1990pj,Hatsuda:1990zj,Adami:1991js,Schafer:1993xt,Saito:1994tq} of the
neutron-proton mass difference in the nuclear medium, while
others gave opposite results~\cite{Meissner:1991if,Fiolhais:1991qf}. Since the Skyrme
model~\cite{Skyrme:1961vq,Skyrme:1962vh}
 treats the properties~\cite{Adkins:1983ya,Adkins:1983hy,Zahed:1986qz}
 {\em and} the 
interactions~\cite{Jackson:1985bn,Kaulfuss,Hajduk,Otofujii}
 of the nucleons on an equal
footing,
it is interesting to study what this 
model has to say for possible  medium-modifications of this quantity.
Note also that the neutron-proton mass difference is composed of a strong and
an electromagnetic contribution of comparable size, which might be affected
differently in a dense medium. 

In the present work we evaluate isospin-breaking effects 
in the baryonic sector of the Skyrme model by incorporating
the influence of the surrounding nuclear environment 
on the nucleons as \textit{e.g.} in 
Refs.~\cite{Rakhimov:1996vq,Yakhshiev:2001ht}.
In order to perform our studies we modify
the nonlinear $\sigma$-model Lagrangian, which is  generalized by 
explicit isospin-breaking effects
in the mesonic sector~\cite{Rathske:1988qt},  based on 
the well-known pionic field equations
in the nuclear medium~\cite{Ericsonbook}.
Furthermore, the usual  Skyrme term is added 
in order to stabilize the solitons of
the model.

The paper is organized as follows.
In Sec.~\ref{sec:mod-lag} we discuss our model Lagrangian. 
Section~\ref{sec:minimization} is devoted to the study of the field equations
to be solved. After an explanation of the quantization procedure
and the derivation of  
the strong part of the density-dependent neutron-proton mass difference
in Sec.~\ref{sec:quantization}, we will present the 
electromagnetic (EM) part of that
mass difference in terms of electromagnetic form factors 
in Sec.~\ref{sec:EMdif}.
The input values of the model parameters used in this work 
are discussed in
Sec.~\ref{sec:parameters}. 
Section~\ref{sec:results} is devoted to the presentation
and discussion of the results. Finally,  in Sec.~\ref{sec:summary} our
conclusions are summarized and an outlook for further studies is given.

\section{Medium-modified Lagrangian}
\label{sec:mod-lag}

We start from a generalized Lagrangian which incorporates
an explicit isospin breaking term  in the mesonic sector
\begin{eqnarray}
{\cal L}&=&{\cal L}_2+{\cal L}_4+{\cal L}_{{\rm g}\chi {\rm SB}}\,,\label{lb}\\
{\cal L}_2&=&-\dsf{F_\pi^2}{16}\,\Tr\, L_\mu L^\mu\,,\\
{\cal L}_4&=&\dsf{1}{32e^2}\,\Tr\,[L_\mu,L_\nu]^2\,,\\
{\cal L}_{{\rm g}\chi {\rm SB}}&=& -\dsf{F_\pi^2}{16}\left\{\Tr\,(U-1){\cal M}_+(U^\dagger-1)
{\cal M}_+\right.\nonumber\\
&&\left.-\Tr\,(U-1){\cal M}_-\tau_3 (U^\dagger-1) {\cal M}_- \tau_3\right\}\,,
\label{LRathske}
\end{eqnarray}
where $L_\mu=U^\dagger\partial_\mu U$. 
The Einstein summation convention for repeated Greek or Latin 
indices is adopted, unless stated otherwise.
${\cal L}_2$ is the usual Lagrangian of the nonlinear sigma model, 
${\cal L}_4$ is Skyrme's stabilizing fourth order term.
${\cal L}_{{\rm g}\chi {\rm SB}}$
is a generalized 
pion-mass term  proposed by Rathske~\cite{Rathske:1988qt} which
includes  explicitly isospin-breaking effects in the second term.
Note that the latter has the same structure as the strong
isospin-breaking fourth-order Lagrangian 
of  chiral perturbation theory, see the term proportional to the
low-energy constant 
$L_7$, Eq.~(6.16) of 
Ref.~\cite{Gasser:1984gg}.~\footnote{This is manifest when the 
isospin-breaking term is rewritten as in Eq.\,\re{lme1}. Because of 
$U U^\dagger = 1$, Rathske's first term, the pion-mass term, is compatible
with
the usual choice of explicit chiral symmetry breaking 
proportional to $\Tr( U+U^\dagger -2)$ \cite{Adkins:1983hy}.}
Skyrme models with isospin-breaking terms proportional to the
$\rho$-$\omega$ meson-mixing, as {\textit e.g.} discussed in 
Refs.~\cite{Epele:1988ak,Jain:1989kn}, are actually of the same order in
the chiral expansion, but more involved because of the additional meson degrees
of freedom.
The  parameter $F_\pi$ is the pion decay constant, 
while  $e$ is the dimensionless 
Skyrme constant. The SU(2) matrix
$U(t,\vec{r})=\exp\left\{{2i}\vec{\tau}\cdot\vec{\Phi}(t,\vec{r})/F_\pi\right\}$
represents the isotriplet of pion fields $\vec{\Phi}$. Their masses,
$m_{\pi^\pm}$ and $m_{\pi^0}$,  enter the Lagrangian in the
following combinations
\begin{eqnarray}
{\cal M}_\pm&=&\sqrt{({m_{\pi^\pm}^2\pm m_{\pi^0}^2})/{2}}\,.
\label{Mplusminus}
\end{eqnarray}

The medium-modified Lagrangian, 
which takes into account the influence of the surrounding
nuclear environment on the pion fields, 
can be set up by following the steps presented in
Ref.~\cite{Rakhimov:1996vq}. It has the form\footnote{{}From now on an asterix
indicates  a density-dependent quantity.}
\begin{eqnarray}
{\cal L}^*&=&
 {\cal L}_2+{\cal L}_4+{\cal L}_{2,{\rm g}\chi {\rm SB}}^*\,,\label{lmb}\\
{\cal L}_{2,{\rm g}\chi {\rm SB}}^*&=&-\dsf{F_\pi^2}{32}\Big\{
-\left(\chi_{\rm p}^\pm+\chi_{\rm p}^0\right)\,\Tr\,\vec{\nabla} 
U\cdot\vec{\nabla} U^\dagger
\nonumber\\
&&+\left(\chi_{\rm p}^\pm-\chi_{\rm p}^0\right)\,
   \Tr\,\vec{\nabla} U\tau_3 \cdot\vec{\nabla} U^\dagger\tau_3
\nonumber\\
&&+\left(2{\cal M}_+^2+\chi_{\rm s}^\pm+\chi_{\rm s}^0\right)
\Tr(U-1)(U^\dagger-1)\nonumber\\
&&-\left(2{\cal M}_-^2+\chi_{\rm s}^\pm-\chi_{\rm s}^0\right)
\Tr(U-1)\tau_3(U^\dagger-1)\tau_3\Big\}.\nonumber
\end{eqnarray}
Here, $\chi_{\rm s,p}^\pm$ and  $\chi_{\rm s,p}^0$ are 
functionals of the low-energy S- and P-wave pion-nucleon scattering 
lengths and volumes, $b_{0,1}$ and $c_{0,1}$, respectively~\cite{Ericsonbook}.  
They express  the influence of the medium
on the charged and neutral 
pion fields, $\pi^\pm$ and $\pi^0$. Furthermore, they  depend on the 
densities of the neutron and proton distributions, 
$\rho_{\rm n}$ and $\rho_{\rm p}$, of the environment. 
One can check that the Lagrangian~\re{lmb} -- in the linear approximation --  
generates the well-known field equations
$(\partial_\mu\partial^\mu+m_{\pi^{\pm,0}}^2
  +\hat\Pi^{\pm,0})\pi^{\pm,0}=0$~\cite{Ericsonbook},
where the polarization operators $\hat\Pi^{\pm,0}$
have the schematic form 
\begin{equation}
\hat\Pi^{\pm,0}=\chi_{\rm s}^{\pm,0}+\vec{\nabla}\chi_{\rm p}^{\pm,0}\cdot
    \vec{\nabla}\,.
\label{Pol-op}
\end{equation}

The medium-modified Lagrangian~\re{lmb} can be rewritten 
in the following way
\begin{eqnarray}
{\cal L}^*&\equiv&{\cal L}_2^*+{\cal L}_4+{\cal L}_{\chi {\rm SB}}^*
+\Delta{\cal L}^*\,,\label{lmb1}\\
{\cal L}_2^*&=&\dsf{F_\pi^2}{16}\Big\{\Tr\partial_0U\partial_0U^\dagger
\!-\!\left(1\!-\!\chi_{\rm p}^0\right)\Tr\vec{\nabla} U\cdot\vec{\nabla} U^\dagger
 \Big\}\,,\\
{\cal L}_{\chi {\rm SB}}^*&=&-\dsf{F_\pi^2 m_{\pi^0}^2}{16}
\left(1+\frac{\chi_{\rm s}^0}{m_{\pi^0}^{2}}\right)\Tr(U\!-\!1)(U^\dagger\!-\!1)\,,\\
\Delta{\cal L}^*&=&\dsf{F_\pi^2}{32}\Big\{
\left(\chi_{\rm p}^\pm-\chi_{\rm p}^0\right)\,
\left[\,\Tr\,(\tau_i\vec{\nabla} U)\cdot
\Tr(\tau_i\vec{\nabla} U^\dagger)\right.\nonumber\\
&&\qquad\left.\mbox{}-\Tr\,(\tau_3\vec{\nabla} U)\cdot
\Tr(\tau_3\vec{\nabla} U^\dagger)\right]
\nonumber\\
&&\quad\ -\left(2{\cal M}_-^2+\chi_{\rm s}^\pm-\chi_{\rm s}^0\right)
\left[\Tr\,(\tau_iU)\,\Tr(\tau_iU^\dagger)\right.\nonumber\\
&&\qquad\left.\mbox{}-\Tr\,(\tau_3U)\,\Tr(\tau_3U^\dagger)\right]\Big\}\,,
\label{lme1}
\end{eqnarray}
where ${\cal L}_{\chi {\rm SB}}^*$ is the chiral-symmetry-breaking term and 
$\Delta{\cal L}^*$ is the isospin-breaking term, separately.
Note that in the absence of $\Delta{\cal L}^*$
the Lagrangian~\re{lmb1} reduces to the 
medium-modified Lagrangian presented in Ref.~\cite{Rakhimov:1996vq}. 

In general,  the strong isospin breaking in the baryonic sector 
is generated by two effects: 
an explicit isospin-breaking
in the mesonic sector (in the Lagrangian 
this term is proportional to ${\cal M}_-^2$) {\em and}
the isospin-dependent influence of the medium  on each of the pion fields,
$\chi_{\rm p}^\pm-\chi_{\rm p}^0$ and $\chi_{\rm s}^\pm-\chi_{\rm s}^0$.
For simplicity and in order to concentrate on the effects due to explicit 
isospin breaking in the mesonic sector, we will
consider from now on only isospin-symmetric nuclear matter:
\begin{eqnarray}
\chi_{\rm p}^\pm&=&\chi_{\rm p}^0\equiv\chi_{\rm p}=
\dsf{4\pi c_0\rho/\eta}{1+4g^\prime\pi c_0\rho/\eta}\,,\nonumber\\
\chi_{\rm s}^\pm&=&\chi_{\rm s}^0\equiv\chi_{\rm s}
 =-4\pi\eta (b_0)_{\rm eff}\rho\,.
\label{mf}
\end{eqnarray}
Here $\eta=1+m_{\pi^0}/m_{\rm N}$ is a kinematical factor, $m_{\rm N}=938$~MeV is 
the nucleon mass,
$g^\prime$ is a correlation parameter,
$c_0$ is the P-wave isoscalar pion-nucleon scattering volume, 
$(b_0)_{\rm eff}$ is the corresponding {\em effective} S-wave isoscalar scattering
length in nuclear matter, 
and $\rho$ is the density 
of the surrounding medium~\cite{Ericsonbook}.
Moreover, we will limit the investigation  to 
homogenous nuclear matter, $\rho_{\rm p}=\rho_{\rm n}=\rho/2={\rm const}$, 
and to the
evaluation of  the neutron-proton mass difference in  nuclear matter, 
$\Delta m_{\rm np}^*=m_{\rm n}^*-m_{\rm p}^*$.

To finish this section, we remark that the approach used here 
should be distinguished
from the method used in 
Refs.~\cite{Meissner:1988wj,Meissner:1988fc,Meissner:1989bd}. 
In these papers, all meson properties like the
masses, decay constants, etc.\ were assumed to be density-dependent and taken 
from a microscopic {\em model} of the meson dynamics. 
Here, the
density-dependence is parameterized in terms of phenomenological input, more
precisely  the effective scattering length $(b_0)_{\rm eff}$ and volume $c_0$.

The Lagrangian \re{lmb} is of course only one representative of a larger
class of Skyrme-type models with {\textit e.g.} 
higher-order terms or vector-meson
degrees of freedom  which, rooted in a hadronic language, do allow
for a simultaneous description
of nucleon properties and interactions. However, the present choice with
its scale-independent dimensionless Skyrme parameter $e$ can be
considered as the simplest, but also as a generic one, with a minimal
number of derivatives and model-parameters,
that incorporates explicitly strong isospin breaking, 
stable solitons, and medium modifications.

\section{Classical solitonic solutions}
\label{sec:minimization}

Due to the isospin-breaking term of the Lagrangian~\re{lmb1}, the
third component of the isovector field is
singled out
in the field equations\footnote{The explicit form
of Eq.\,\re{fieq} can be found
in Ref.~\cite{Rathske:1988qt} for the free space case. Note that the
Skyrme term is omitted there.}:
\begin{equation}
\partial_\mu\dsf{\partial{\cal L}^*}{\partial\partial_\mu\vec{\Phi}}-
\dsf{\partial{\cal L}^*}{\partial\vec{\Phi}}=0\,.
\label{fieq}
\end{equation}
Treating the (strong) isospin-breaking as a small perturbation, we can
distinguish between a  non-perturbed baryonic 
background with $m_{\pi^\pm}^2 = m_{\pi^0}^2$
and perturbative  contributions proportional to the very tiny {\em
 strong-interaction} induced part of 
$m_{\pi^\pm}^2- m_{\pi^0}^2\ne 0$.
The 
non-perturbed system has the usual time-indepen\-dent classical 
(baryon number $B=1$) 
solutions $\vec{\Phi}(t=0,\vec{r})=\vec{\phi}(\vec{r})$ of the 
isospin-symmetric Lagrangian,
whereas a per\-turbation a\-round 
the classical background $\vec{\phi}(\vec{r})$
can be absorbed {\textit e.g.} by a time-\-de\-pendent modification. 

We follow Rathske \cite{Rathske:1988qt} by using for this purpose 
the separation ansatz
\begin{equation}
\vec{\Phi}(t,\vec{r})={\cal T}(t)\,\vec{\phi}(\vec{r})\,, 
\label{t-anz}
\end{equation}
where a manifest  isospin-breaking  
(with respect to the 3rd isospin-axis) is encoded in the time-dependent 
iso-rotation 
matrix ${\cal T}$: 
\begin{equation}
{\cal T}(t)=\left(\begin{array}{rrr}
\cos a^* t&\sin a^* t&0\\
-\sin a^* t&\cos a^* t& 0\\
0&0&1\end{array}\right)\,.
\label{Tmatrix}
\end{equation}
Here $a^*$ is a density-dependent global constant\footnote{Note that a sign 
change of $t$ can be compensated by a sign change of the 
angular velocity  $a^*$.}
which would vanish in the absence
of the small perturbation $\Delta m_{\pi}=m_{\pi^\pm}-m_{\pi^0}$, since
then the system would reduce to the non-perturbed configuration
$\ds\lim_{\Delta m_{\pi}\rightarrow 0}{\cal L}^*={\cal L}_{\rm NP}^*$.

The ansatz \re{t-anz} together with  \re{Tmatrix}, when inserted into the
Lagrangian \re{lmb1} and after spatial integration, allows to cancel
the non-perturbative shift (IB-NP) proportional to $m_{\pi^\pm}^2-m_{\pi^0}^2$
that results from inserting the non-perturbative
(NP) background  $\vec{\phi}(\vec{r})$ of the isospin-symmetric case 
into the isospin-breaking  
part (IB) of the
Lagrange function against the  isospin-breaking
term (IB-{$\cal T$}) that results 
from the time-dependent rotation by the $\cal T$ matrix:
\begin{eqnarray}
&&L^* =\int {\cal L}^*\, {\rm d}^3 {r}
=-M_{\rm NP}^* -{\cal M}_-^2\Lambda_-^* +\dsf{a^{*2}}{2}\Lambda^*\nonumber\\
&&\equiv
\int{\cal L}_{\rm NP}^*\,{\rm d}^3{r}
+\int{\cal L}_{{\rm IB}-{\rm NP}}^*\,{\rm d}^3{r}
+\int{\cal L}_{{\rm IB}-{\cal T}}^*\,{\rm d}^3{r}\,.
\label{lag1}
\end{eqnarray}
Here $M_{\rm NP}^*[\vec{\phi}]$ is the isospin-symmetric 
mass of the non-perturbed configuration. The moment-of-inertia type 
quantities $\Lambda^*=\Lambda^*[\vec{\phi}]$ and
$\Lambda_-^*=\Lambda_-^*[\vec{\phi}]$, belonging to the $\cal T$-rotation 
and the non-perturbative 
shift, respectively, are functionals of the
pion fields. The global parameter  $a^*$ serves  here as a {\em constraint} 
parameter, whereas in Ref.\,\cite{Rathske:1988qt} 
-- under the neglect of 
stabilizing higher-order terms --  the ansatz   \re{t-anz} 
also extremized the remaining action.
Thus the value of $a^{*2}$ is fixed here 
by the following condition which implies
the cancellation between the IB-NP and IB-${\cal T}$ parts in the Lagrange 
function:
\begin{equation}
a^{*2}=2{\cal M}_-^2\dsf{\Lambda_-^*}{\Lambda^*}\,.
\label{cond}
\end{equation}
As the iso-rotation matrix $\cal T$ is chosen in such a way that
-- at the classical level and after spatial integration -- 
the isospin-breaking is rotated out of  the system,
the {\em classical} field equations  \re{fieq} 
effectively reduce to the non-perturbed isospin-symmetric case:
\begin{eqnarray}
\partial_\mu\dsf{\partial{\cal L}^*}{\partial\partial_\mu\vec{\Phi}}-
\dsf{\partial{\cal L}^*}{\partial\vec{\Phi}}=0
&\quad \Rightarrow\quad&
\dsf{\delta M_{\rm NP}^*[\vec{\phi}]}{\delta\vec{\phi}}=0\,.
\label{fieq2}
\end{eqnarray}

In general, one should consider non-spherical time-inde\-pen\-dent
field configurations $\vec{\phi}(\vec{r})$  if
the density gradients of the surrounding nuclear
environment are large. In this case,
the static field configuration 
acquires a $\theta$-dependence~\cite{Yakhshiev:2001ht}, 
\textit{i.e.} $\vec{\phi}$ = $\vec{\phi}(|\vec{r}|,\theta)$, even
 if isospin-breaking  terms
are absent. The above presented ``rotation'' procedure has the
advantage that it remains applicable for very non-uniform 
density-profile-dependent cases,  in other words, for finite nuclei. 
But as stated in 
Sec.~\ref{sec:mod-lag}, only the case of an
isospin-symmetric homogeneous
nuclear environment with a constant density is considered here. Thus
the use of a spherically symmetric static  configuration  
$\vec{\phi}(\vec{r})$ = $({\vec{r}}/r) {F_\pi} F(r)/2$
is still appropriate. Under this  {\em hedgehog} ansatz
the terms defined in~\re{lag1} are given by
\begin{eqnarray}
M_{\rm NP}^*&=&\pi\int\limits_0^\infty\Big\{
\dsf{F_\pi^2}{2}\left(1-\chi_{\rm p}\right)\left(F_r^2 +\frac{2\, \sF^2}{r^2}\right)
\nonumber\\
&&
+\dsf{2}{e^2}\left(2F_r^2+\frac{\sF^2}{r^2}\right)\frac{\sF^2}{r^2}\nonumber\\
&&+F_\pi^2\left( m_{\pi^0}^2 + {\chi_{\rm s}}\right)\left(1-\cos\!F\right)
\Big\} r^2\,{\rm d}r\,,
\label{clEm0}\\
\Lambda^*&=&\dsf{2\pi}{3}\int\limits_0^\infty \left\{
F_\pi^2+\dsf{4}{e^2}
\left(F_r^2+\frac{\sF^2}{r^2}\right)\right\} \sF^2 \,r^2\,{\rm d}r\,,
\label{clEm}\\
\Lambda_-^*&=&\dsf{2\pi}{3}F_\pi^2\int\limits_0^\infty \sF^2\, 
   r^2\,{\rm d}r\,,
\label{clEe}
\end{eqnarray}
where $F_r\equiv {\rm d} F(r)/{\rm d}r$. Also the  abbreviation 
$\sF\equiv\sin F(r)$ is used here.

The corresponding field equations~\re{fieq2} do not explicitly depend on the
isospin index  and therefore reduce to the simple radial form 
that is 
used in this work to determine the profile function $F(r)$:
\begin{eqnarray}
&&\dsf{F_\pi^2}{2}\left(1-\chi_{\rm p}\right)
\left(F_{rr}+\dsf{2}{r}F_r-\dsf{\sin2F}{r^2}\right)\nonumber\\
&&\mbox{}+\dsf{2}{e^2}\left(\frac{2\sin^2 F}{r^2} F_{rr}+\dsf{\sin2F}{r^2}
\left(F_r^2-\frac{\sin^2 F}{r^2}\right)\right)\nonumber\\
&&\mbox{}-\dsf{F_\pi^2}{2}\left(m_{\pi^0}^2+{\chi_{\rm s}}\right)\sin\!F=0\,.
\label{NP-eq}
\end{eqnarray}
Here $F_{rr}$ stands for
the second derivative of the profile function $F(r)$ with respect to the
radial coordinate $r$.

If the Skyrme
term were omitted,  the condition~\re{cond} would reduce to 
$a^{*2}=m_{\pi^\pm}^2-m_{\pi^0}^2$  and the results of the work~\cite{Rathske:1988qt}
would be reproduced in a natural way.
However in this case, due to the then 
profile-independent 
value of the parameter $a^{*}$,
the density dependence of the isospin-breaking
effects would be lost\footnote{The isospin-breaking effects resulting from
the  pion-mass variation in the nuclear medium
are beyond the scope of the present model.}.
{}From this point of view, the presence of 
a higher-order term in the Lagrangian is crucial,
not only for its stabilization role, but also
for the description of the {\em density-dependence} 
of the  isospin-breaking in the  nuclear environment.
Of course, 
there is a model-dependence due to the nature of the 
stabilizing term. We reiterate that the Skyrme term 
is the most simplest choice,
with
the least number of derivatives and model-constants, that guarantees the
existence of stable  baryons. In fact, the chiral order of the Skyrme term
is the same as the one of the strong isospin-breaking term of chiral
perturbation theory.

To finish  this section, let us state the expectation that
the above mentioned shift induced by the 
isospin-breaking terms and the compensating
isospin-breaking rotation 
will be manifest  in the quantized theory, 
although they are constrained to cancel 
each other at the classical level.

\section{Quantization and the strong part of 
$\mathbf{\Delta m_{\rm np}^*}$}
\label{sec:quantization}

The standard quantization procedure of the Skyrme mo\-del~\cite{Adkins:1983ya} 
requires  time-dependent rotations in space, ${\cal R}(t)$,  and 
isospin-space, 
${\cal I}(t)$: 
\begin{eqnarray}
U(t,\vec{r})&=&\exp\left(i\frac{2}{F_\pi} \vec{\tau}\cdot\
\vec{\Phi}(t,\vec{r})\right)\nonumber \\
\hookrightarrow
U^\prime(t,{\vec{r}}^{\prime})
&=&\exp\left(i\frac{2}{F_\pi} \vec{\tau}
\cdot {\cal I}(t)
\vec{\Phi}\left(t,{\cal R}^{-1}(t)\vec{r}\right)\right)\,,
\end{eqnarray}
where $\vec{\Phi}(t,\vec{r})$ is defined in  Eq.~\re{t-anz}.
The rotations ${\cal R}(t)$  and 
iso-rotations ${\cal I}(t)$ correspond to standard
collective zero-energy modes of the classical soliton,
whereas the matrix ${\cal T}$ of Eq.\,\re{t-anz} describes a
{\em constrained} rotational mode with respect to the classical soliton.
The SO(3) matrices, ${\cal I}(t)$ and ${\cal R}(t)$, satisfy the conditions
\begin{equation}
\dot{{\cal I}}_{ik}(t){\cal I}_{kj}^{-1}(t)=\varepsilon_{ijl}\omega_l\,,
\quad
\dot{{\cal R}}_{ik}^{-1}(t){\cal R}_{kj}(t)=-\varepsilon_{ijl}\Omega_l\,,
\end{equation}
where $\omega_l$ and $\Omega_l$ are the 
angular velocities of the isospin  and spatial rotations, respectively,
and the `dot' symbolizes a time derivative.
Under these rotations and  the constraint \re{cond} 
the corresponding  Lagrange function
${L}^*=\int  {\cal L}^*\,{\rm }d^3 {r}$ takes the form
\begin{eqnarray}
{L}^*
&=&-M_{\rm NP}^*+\dsf{\Lambda^*}{2}
\left[ {\left( \vec{\omega} -\vec{\Omega}\right)}^2
+2a^*\left(\omega _3 - \Omega _3 \right)\right].
\label{lagquan}
\end{eqnarray}
Note that the third components of the angular velocities couple 
to the constrained angular velocity $a^*$.
Finally, the usual Legendre transformation utilizing the 
canonical-con\-jugated operators
$\hat T_i=\partial_{\omega_i}{L^*}\!$,
$\hat J_i=\partial_{\Omega_i}{L^*}$
of $\omega_i$ and $\Omega_i$, respectively,
leads to  the quantum Hamiltonian:
\begin{eqnarray}
{\hat H^*}  &=&M_{\rm NP}^*
+\dsf{\hat{\vec{T_1}}^2}{2\Lambda^*} 
+\dsf{\hat{\vec{T_2}}^2}{2\Lambda^*}
+\dsf{\left(\hat{\vec{T_3}}-\Lambda^* a^*\right)^2}{2\Lambda^*} \nonumber\\
&=&M_{\rm NP}^*+{\cal M}_-^2\Lambda_-^*
+\dsf{\hat{\vec{T}}^2}{2\Lambda^*}-a^*\hat T_3\,.
\label{Ham}
\end{eqnarray}
There
does not exist a canonical-conjugated momentum of the angular velocity 
$a^*$ since this is a
constrained quantity.
The mass-shift term due to  the explicit isospin-breaking, 
the second term of Eq.\,\re{Ham},  
reappears  -- compared with Eq.\,\re{lag1} or Eq.\,\re{lagquan} --
after applying  the constraint \re{cond} to the induced 
term $(\Lambda^*/2){a^*}^2$. 
Moreover, note
that the isospin breaking in the mesonic sector 
is  manifest at the quantum level by the coefficient of  
the third component of the isospin
operator, see the fourth term of Eq.\,\re{Ham}.
Consequently, by evaluating the quantum
Hamiltonian~\re{Ham} between appropriate nucleon states one can
isolate the strong part of the neutron-proton mass difference in the
nuclear medium as~\footnote{Note 
that the sign of $a^*$ is now fixed by
the known sign of the strong nucleon-proton mass difference in free space.}
\begin{equation}
m_{\rm n}^{*(\rm strong)}-m_{\rm p}^{*(\rm strong)}
=\Delta m_{\rm np}^{*(\rm strong)}=a^*\,.
\label{mn-strong}
\end{equation}
This term is enhanced in comparison to the tiny mass 
shift ${\cal M}_-^2\Lambda_-^*$, since the latter is quadratic and not linear 
in the small parameter $a^*$.

Finally, 
let us discuss the scaling  under the expansion in the number of colors $N_c$ 
of QCD, the underlying theory of strong interactions. From the viewpoint of 
the non-relativistic quark model, the mass difference of a large $N_c$
neutron (with $(N_c+1)/2$ $d$-quarks and $(N_c-1)/2$ $u$-quarks)  and 
a large $N_c$ proton (with $(N_c+1)/2$ $u$-quarks and $(N_c-1)/2$ $d$-quarks) is
expected to scale as  $m_d-m_u\sim N_c^0$. Let us compare this with our
Eq.\,\re{mn-strong}: As usual, we start with the assumption that
$F_\pi \sim 1/e \sim \sqrt{N_c}$, such
that the soliton mass $M_{\rm NP}^*$, the moment-of-inertia $\Lambda^*$ 
and also  $\Lambda_-^*$ scale 
as ${\cal  O}(N_c^1)$. Assuming that ${\cal M}_-$ scales as
${\cal M}_+$, namely as ${\cal O}(N_c^0)$,~\footnote{Note 
that the $N_c$ scaling of the
SU(3)  low-energy constant $L_7\sim (1/48) F_\pi^2/m_{\eta'}^2\sim {\cal
  O}(N_c^2)$~\cite{Gasser:1984gg}  would actually imply 
that  ${\cal M}_-^2$
should scale as ${\cal O}(N_c^1)$ (see \re{LRathske}), 
since  the $\eta'$ would become a Goldstone
boson in the limit $N_c\to \infty$ ({\textit i.e.}, the $\eta'$ mass scales as 
$m_{\eta'}\sim \sqrt{1/N}$). For $N_c=3$, however, the prefactor is
very tiny, such that the above assumed ${\cal O}(N_c^0)$ behavior of 
${\cal M}_-^2 =( m^2_{\pi^\pm}-m^2_{\pi^0})/2$ is the more natural choice for
the strong pion mass difference.} 
the constrained constant $a^*$
(see Eq.\,\re{cond})
and the strong neutron-proton mass difference \re{mn-strong} is
of the same order ${\cal  O}(N_c^0)$. Therefore, the $N_c$ scaling of 
Eq.\,\re{Ham} is as it should be:
the non-perturbative isospin-symmetric mass
term $M_{\rm NP}^*$  and the non-perturbative (via isospin-breaking induced)  
mass shift 
${\cal M}_-^2\Lambda_-^*$  scale both as
${\cal  O}(N_c^1)$; 
the quantization term 
${\hat{\vec{T}}^2}/{(2\Lambda^*)}$ scales as ${\cal  O}(N_c^{-1})$, as
expected
for the quantization of  zero-modes of a 
soliton~\cite{Adkins:1983ya};
finally
the isospin-breaking term $a^*\hat T_3$ resulting from
the  constrained iso-rotation (with the enhancement by $\Lambda^*\sim N_c$ 
relative to the quantized isospin, see
the first line of Eq.\,\re{Ham}) scales as
${\cal  O}(N_c^0)$.

\section{Electromagnetic form factors and the electromagnetic part 
of $\mathbf{\Delta m_{\rm np}^{*}}$}
\label{sec:EMdif}

The electric (E) and magnetic (M) form factors of the nucleon are defined
through the expressions
\begin{eqnarray}
G_{\rm E}^* ({\vec{q}}^2)&=&{\textstyle\frac{1}{2}}
\int {\rm d}^3 {r} \, e^{i{\vec{q}}\cdot{\vec{r}}}j^0(r)~,\nonumber\\
G_{\rm M}^* ({\vec{q}}^2)&=&{\textstyle\frac{1}{2}}\,
m_{\rm N}\int {\rm d}^3{r} \, e^{i{\vec{q}}\cdot{\vec{r}}}[\vec{r}\times 
\vec{j}(r)]~,
\label{ffdef}
\end{eqnarray}
where ${\vec{q}}^2$ is the squared  momentum transfer. 
Furthermore, 
$j^0$ and $\vec{j}$ correspond
to the time and space components of the properly normalized sum
of the baryonic current $B_\mu^*$
and the third component of the isovector current $\vec{V}_\mu^*$ of the Skyrme
model, \textit{i.e.}
\begin{eqnarray}
B_\mu^* &=& \dsf{1}{24\pi^2}\, \varepsilon_{\mu\nu\alpha\beta}
\, \Tr \,L^{\nu}L^{\alpha}L^{\beta}\,,\\
V_{\mu}^{(3)*} &=&\dsf{i}{16}\,\Tr\,\left\{\tau_3\left(-F_{\pi}^2\,C_{\mu}
L_{\mu}+\dsf{1}{{e}^{2}}\,\bigl[L^\nu,[L_{\mu},L_{\nu}]\bigr]
\right)\right\}\nonumber\\
&&+\,\,(\,L\,\rightarrow\, R\,)\,,
\label{VecCur}\\
C_{\mu}&=&\left\{\begin{array}{ccl}
1 &,\ &\mbox{$\mu=0$}\,,\\
1-\chi_{\rm p}&,\ & \mbox{$\mu=1,2,3$}\,.\\
\end{array}\right.
\label{currents}
\end{eqnarray}
Here $\varepsilon_{\mu\nu\alpha\beta}$ is the totally antisymmetric
tensor in four dimensions and the definition 
$R_\mu=U\partial_\mu U^\dagger$ is used. Note that the index $\mu$ 
in Eq.~\re{VecCur} is not
summed over.
By evaluating these current operators
between appropriate nucleon states, one obtains the 
density-dependent electromagnetic
form factors of the 
nucleon~\cite{Meissner:1988wj,Meissner:1988fc,Meissner:1989bd,Musakhanov:1999ug,Yakhshiev:2002sr}.

Although the isoscalar (S) electric and magnetic form factors of
the nucleon 
\begin{eqnarray}
G_{\rm E}^{{\rm S}*}&=&-\dsf{1}{\pi}\int\limits_0^\infty F_r\,\sin^2\!F\, 
 j_0(qr)\,{\rm d}r\,,\nonumber\\
G_{\rm M}^{{\rm S}*}&=&-\dsf{m_{\rm N}}{\pi\Lambda^*}\int\limits_0^\infty
F_r\,\sin^2\!F\,r^2\,\dsf{j_1(qr)}{qr}\,{\rm d}r\,
 \label{gems}
\end{eqnarray}
do  not explicitly depend on the density,
there still exists an {\em implicit} medium modification  caused by the density
dependence of the profile function $F$. This holds also for the
isovector (V) electric form factor 
\begin{eqnarray}
G_{\rm E}^{{\rm V}*}&=&\dsf{\pi}{3\Lambda^*} \int\limits_0^\infty
\left\{F_\pi^2+\dsf{4}{e^2}
\left(F_r^2+\frac{\sin^2 F}{r^2}\right)\right\}
\nonumber\\&&\qquad\times
\sin^2\!F\,r^2\, j_0(qr)\,{\rm d}r\,.
\label{gev}
\end{eqnarray}
The isovector magnetic form factor, however, has both explicit and
implicit density dependences:
\begin{eqnarray}
G_{\rm M}^{{\rm V}*}&=&\dsf{2\pi{m_{\rm N}}}{3}\int\limits_0^\infty
\left\{F_\pi^2(1-\chi_{\rm p})+\dsf{4}{e^2}
\left(F_r^2+\frac{\sin^2 F} {r^2}\right)\right\}\nonumber\\
&&\qquad\mbox{}\times \sin^2\!F\,r^2\,\dsf{j_1(qr)}{qr}\,{\rm d}r\,.
\label{gmv}
\end{eqnarray}
The medium-dependent form factors of the proton and neutron are defined as 
$G_{\rm E,M}^{\left({\rm p}\atop {\rm n}\right)*}=G_{\rm E,M}^{{\rm S}*}
\pm G_{E,M}^{{\rm V}*}$
with the normalization conditions
$G_{\rm E}^{{\rm p}*}(0)$ = $1$, $G_{\rm E}^{{\rm n}*}(0)$ = $0$, 
$G_{\rm M}^{{\rm p}*}(0)$ = $\mu_{\rm p}^*$, 
$G_{\rm M}^{{\rm n}*}(0)$ = $\mu_{\rm n}^*$,
where $\mu_{\rm p}^*$ and $\mu_{\rm n}^*$ are the in-medium 
magnetic moments of the proton and neutron,
respectively. Note that the magnetic moments are measured in terms of the
nuclear Bohr magneton (abbreviated as n.m.\ in Tab.~\ref{table1})  with the 
{\em free-space} nucleon mass $m_{\rm N}$. Therefore, 
$m_{\rm N}$ and not $m_{\rm N}^*$ is used
in Eqs.~\re{ffdef}, \re{gems} and \re{gmv}.

Finally,
for calculating the electromagnetic part of 
the neutron-proton mass difference,
one can apply the formula
\begin{eqnarray}
\Delta m_{\rm np}^{*(\rm EM)}= -\dsf{4\alpha}{\pi}\int\limits_0^\infty {\rm d}q
\Big\{G_{\rm E}^{{\rm S}*}(\vec{q}^2)G_{\rm E}^{{\rm V}*}(\vec{q}^2)\qquad\nonumber\\
-\dsf{\vec{q}^2}{2m_{\rm N}^2}
G_{\rm M}^{{\rm S}*}(\vec{q}^2)G_{\rm M}^{{\rm V}*}(\vec{q}^2)\Big\}\,,
\label{dm-EM}
\end{eqnarray}
see \textit{e.g.}  Ref.~\cite{Gasser:1982ap}. Here
$\alpha={{\rm e}^2}/{4\pi}\approx 1/137$ is the electromagnetic fine
structure constant (with ${\rm e}$ the elementary charge). Note also that
the right-hand-side of this equation, after Eqs.~\re{gems} to \re{gmv} are
inserted, does not depend on the value of the
nucleon mass and moreover that $\Delta m_{\rm np}^{*(\rm EM)}$ scales 
as ${\cal  O}(N_c^0)$  like $\Delta m_{\rm np}^{*(\rm strong)}$.

\section{Input parameters of the model}
\label{sec:parameters}

As input for the free mass of the neutral pion
we take the PDG-value~\cite{Yao:2006px}:
$m_{\pi^0}=134.977$~MeV.
This choice induces the values $F_{\pi}= 108.11$~MeV  and $e = 4.826$,
if one insists on reproducing the empirical (isospin-averaged) 
masses of the nucleon and delta,
$m_{\rm N} = 938$~MeV and $M_{\Delta} = 1232$~MeV, in free space ($\rho=0$)
and without isospin breaking term (${\cal M}_-=0$). 
The small differences to 
the values $F_{\pi}=108$~MeV  and $e=4.84$ of Ref.~\cite{Adkins:1983hy} 
are caused by the value $m_{\pi^0}=138$~MeV used there.

The mass $m_{\pi^\pm}$  of the charged pions, actually only 
its strong contribution\footnote{The electromagnetic contribution to 
$m_{\pi^\pm}$ is not considered throughout this paper.}
to $\Delta m_\pi=m_{\pi^\pm}-m_{\pi^0}$,
is extracted as a variational
parameter from the fit to
the empirical value $\Delta m_{\rm np}^{(\rm Exp)}=1.29$~MeV in
free space. To do this, we first calculate the electromagnetic part of the 
neutron-proton mass difference in free space
in analogy  to the formula~\re{dm-EM},
$\Delta m_{\rm np}^{(\rm EM)}=-0.68$~MeV. This is  in good agreement with
the result of  Ebrahim and Savci~\cite{Ebrahim:1987mu} obtained long
ago. Model calculations, which include $\pi$, $\rho$ and
$\omega$ mesons, also lead to similar results~\cite{Kaiser:1988bt}.
Moreover, these results also are consistent with the
value $\Delta m_{\rm np}^{(\rm EM)} = (-0.76\pm0.30)$~MeV estimated
from the Cottingham formula in Ref.~\cite{Gasser:1982ap}.
Finally, applying
\begin{equation}
m_{\left({\rm p}\atop {\rm n}\right)}=
m_{\left({\rm p}\atop {\rm n}\right)}^{(\rm strong)}\mp 
   \dsf{1}2\Delta m_{\rm np}^{(\rm EM)}\,,
\end{equation}
we find the following estimate
of the neutron-proton mass
difference due to the strong interactions:
$\Delta m_{\rm np}^{(\rm strong)}=\Delta m_{\rm np}^{(\rm Exp)}-\Delta m_{\rm np}^{(\rm EM)}
= 1.97$~MeV which is fitted in the present model by adjusting
(the strong part of) $\Delta m_\pi$,
\textit{i.e.}
the quantities ${\cal M}_+$ and ${\cal M}_-$ of Eq.~\re{Mplusminus}.
In contrast to the sizable value of $\Delta m_{\rm np}^{(\rm strong)}$ the
pion-mass difference due to strong interactions is very
small\footnote{In Ref.~\cite{Rathske:1988qt}, where the Skyrme term is
absent, this value is even equal to $\approx 0.01$~MeV.}:
$\Delta m_{\pi}^{(\rm strong)}\simeq 0.04$~MeV, compatible
with the results of Ref.~\cite{Gasser:1982ap}.

Finally, the parameters of the pion self-energy are taken from
Ref.~\cite{Ericsonbook}, Table 6.2:
$g^\prime_0=1/3$, 
$c_0=0.21m_\pi^{-3}$, and
$(b_0)_{\rm eff}=-0.024m_\pi^{-1}$. The latter two are based on the tabulated
isospin $1/2$ and $3/2$  S$_{1/2}$,
P$_{1/2}$ and P$_{3/2}$  pion-nucleon scattering lengths
from Ref.~\cite{Hoehler:1983} which are still compatible 
with modern calculations in the framework of chiral perturbation theory,
see {\textit e.g.} Tab.~2 of  Ref.~\cite{Fettes:2000xg}.

\section{Results and discussion}
\label{sec:results}

The medium-dependent effective
masses and magnetic moments of the nucleons are presented in 
Tab.~\ref{table1} for a couple of values of the nuclear density.
As expected, the effective masses decrease with increasing density 
of the nuclear medium.
The most dominant contribution to these changes comes from the 
explicit medium $(1-\chi_{\rm p})$
factor in the expression of the mass functional $M_{\rm NP}^*$ of 
the non-perturbed system~\re{clEm0}.
\begin{table}
\caption{Masses and magnetic moments of the nucleons in 
nuclear matter with a density $\rho=\lambda\rho_0$, where
$\rho_0=0.5m_{\pi^0}^3$ is the saturation density of  ordinary
nuclear matter.} \label{table1}
\begin{tabular}{cccccc}\hline
\noalign{\smallskip}
$\lambda$&$m_{\rm p}^*$~[MeV]\,&$m_{\rm n}^*$~[MeV]\,&$\mu_{\rm p}^*$~[n.m.]\,&$\mu_{\rm n}^*$~[n.m.]\\
\noalign{\smallskip}\hline\noalign{\smallskip}
0    &937.4 &938.7 & 1.99 & $-$1.27\\
0.2\,&861.9 &863.2 & 1.86 & $-$1.13\\
0.4\,&796.6 &797.9 & 1.75 & $-$1.02\\
0.6\,&739.6 &740.9 & 1.65 & $-$0.92\\
0.8\,&689.5 &690.8 & 1.56 & $-$0.83\\
1.0\,&645.2 &646.4 & 1.48 & $-$0.75\\
\noalign{\smallskip}\hline
\end{tabular}
\end{table}

\begin{figure}
\centerline{
\resizebox{0.41\textwidth}{!}{\includegraphics{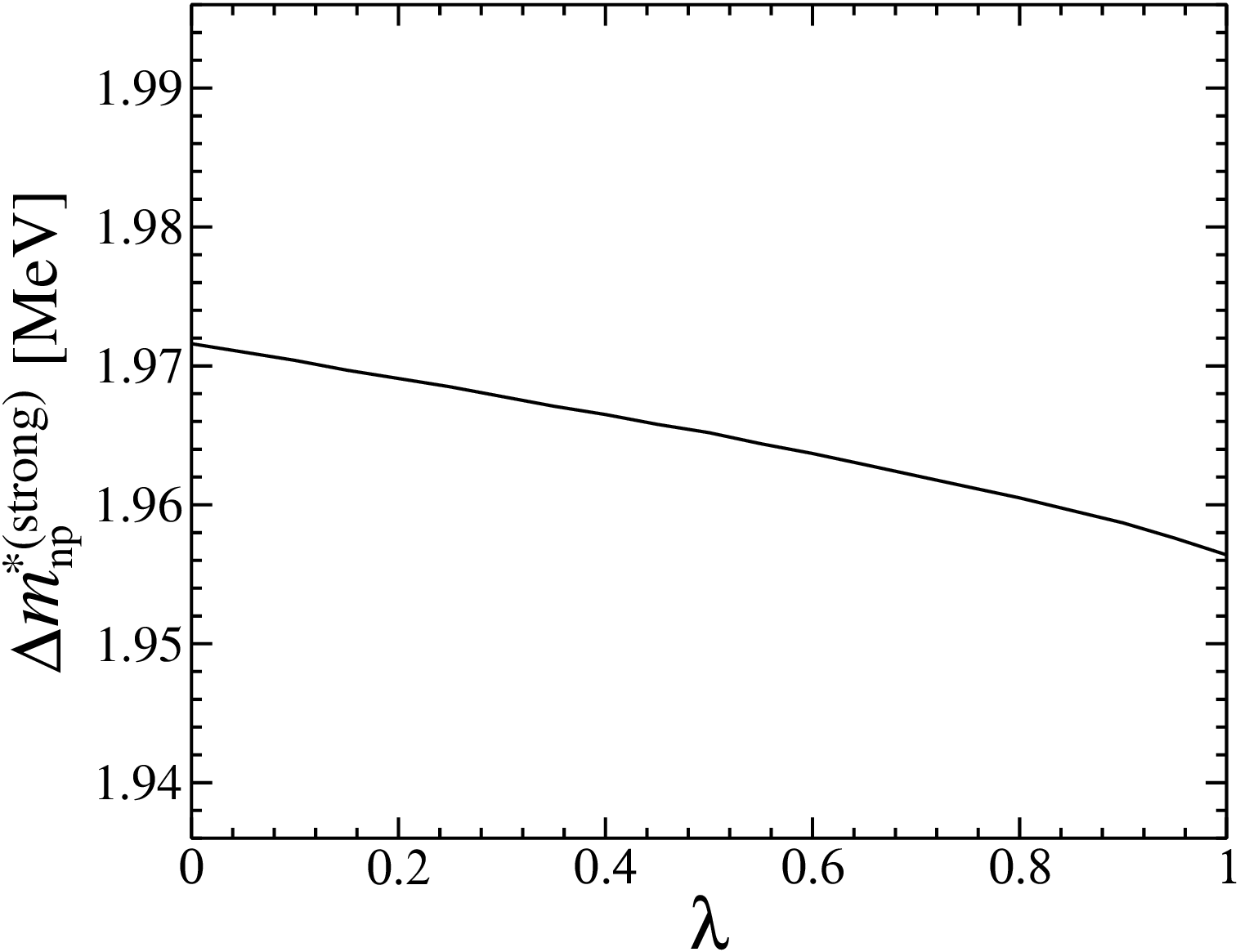}}}
\vspace{0.2cm}

\caption{Density dependence of the strong part of the 
neutron-proton mass difference.
The abscissa represents $\lambda=\rho/\rho_0$,
while the ordinate represents  
$\Delta m_{\rm np}^{*(\rm strong)}$ in units of MeV.}
\label{fig1}
\end{figure}

Of course, it is well established  that the effective mass of the nucleon 
decreases in the medium~\cite{Jeukenne:1976uy,Mahaux85}. 
However, the in-medium behavior of the neutron-proton
mass difference is still under debate. Our results for 
the mass difference due to
the strong interaction, $\Delta m_{\rm np}^{*(\rm strong)}$, and electromagnetic
interaction, $\Delta m_{\rm np}^{*(\rm EM)}$, are presented in Fig.~\ref{fig1} and
Fig.~2, respectively.
\begin{figure}
\centerline{\resizebox{0.41\textwidth}{!}{\includegraphics{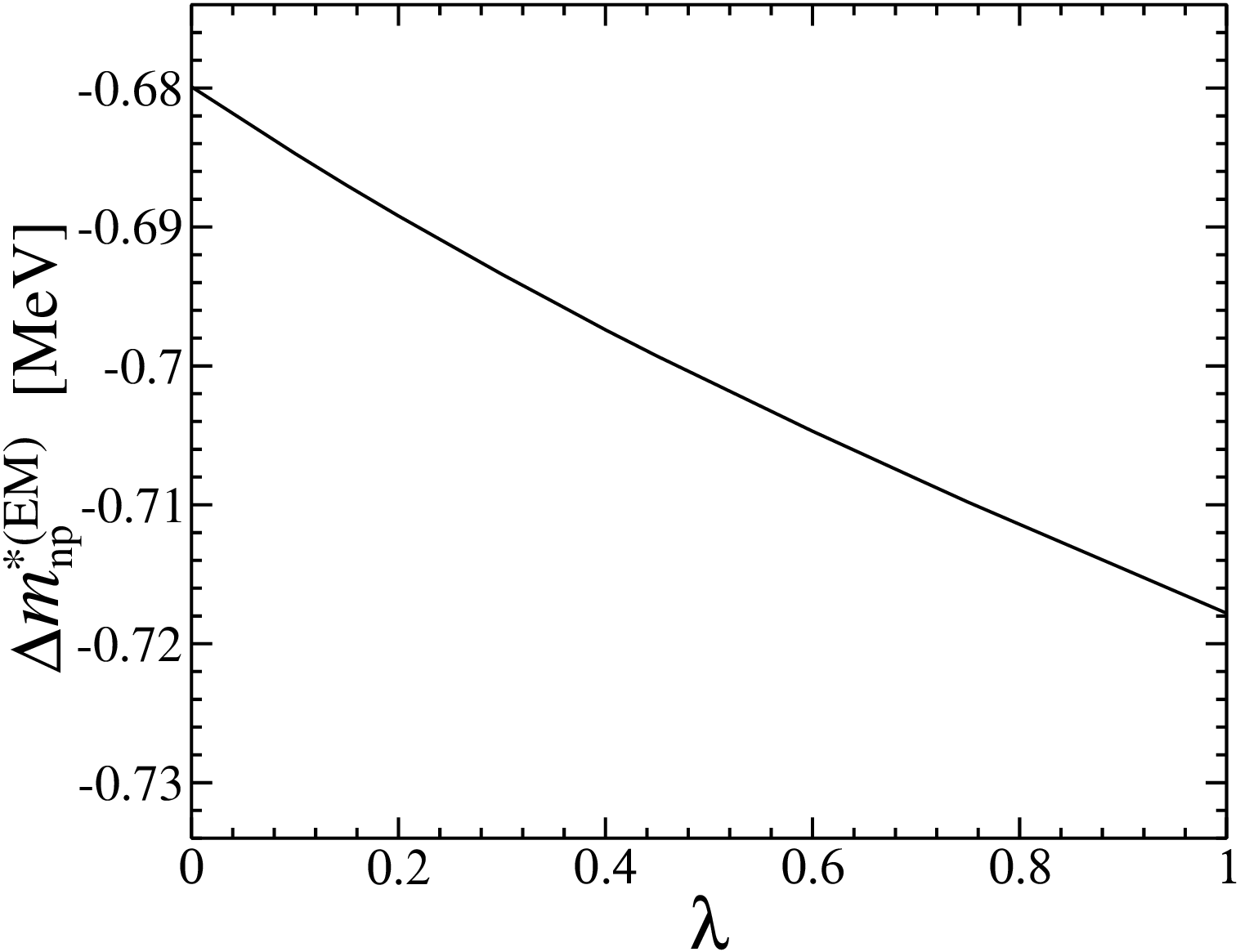}}}
\vspace{0.2cm}      
\caption{Density dependence of the 
electromagnetic part of the neutron-proton
mass difference.
The abscissa represents $\lambda=\rho/\rho_0$,
while the ordinate represents  $\Delta m_{\rm np}^{*(\rm EM)}$ in units of
MeV.}
\label{fig2}
\end{figure}

It can be seen that both parts, the strong and electromagnetic one,
have an almost linear dependence on the medium density and decrease with
increasing the density. Note, however, that the absolute values of the
changes are very small: \textit{i.e.},
the strong part of  $\Delta m_{\rm np}^*$ is almost flat (see
Fig.~\ref{fig1}), while the change in the electromagnetic part  
is slightly more pronounced (see Fig.~\ref{fig2}). 
The difference in their behavior
follows from the fact that $\Delta m_{\rm np}^{*(\rm strong)}$ has no 
{\em explicit}  dependence on the medium functionals (see Eqs.~\re{cond}
and~\re{mn-strong}), whereas  $\Delta m_{\rm np}^{*(\rm EM)}$~\re{dm-EM}
explicitly depends on the P-wave medium functional via the
isovector magnetic form factor~\re{gmv}. However, there is a chance that
the corresponding shifts may be more pronounced
in isospin-asymmetric matter, due to the pertinent
in-medium functionals in the isospin breaking term of the
Lagrangian $\Delta{\cal L}^*$ (see Eq.~\re{lme1})
and the additional dependence of the medium functionals on the
difference of the neutron-proton (distribution)  densities of the
surrounding environment, $\delta\rho=\rho_{\rm n}-\rho_{\rm p}$.

\begin{figure}
\centerline{\resizebox{0.41\textwidth}{!}{\includegraphics{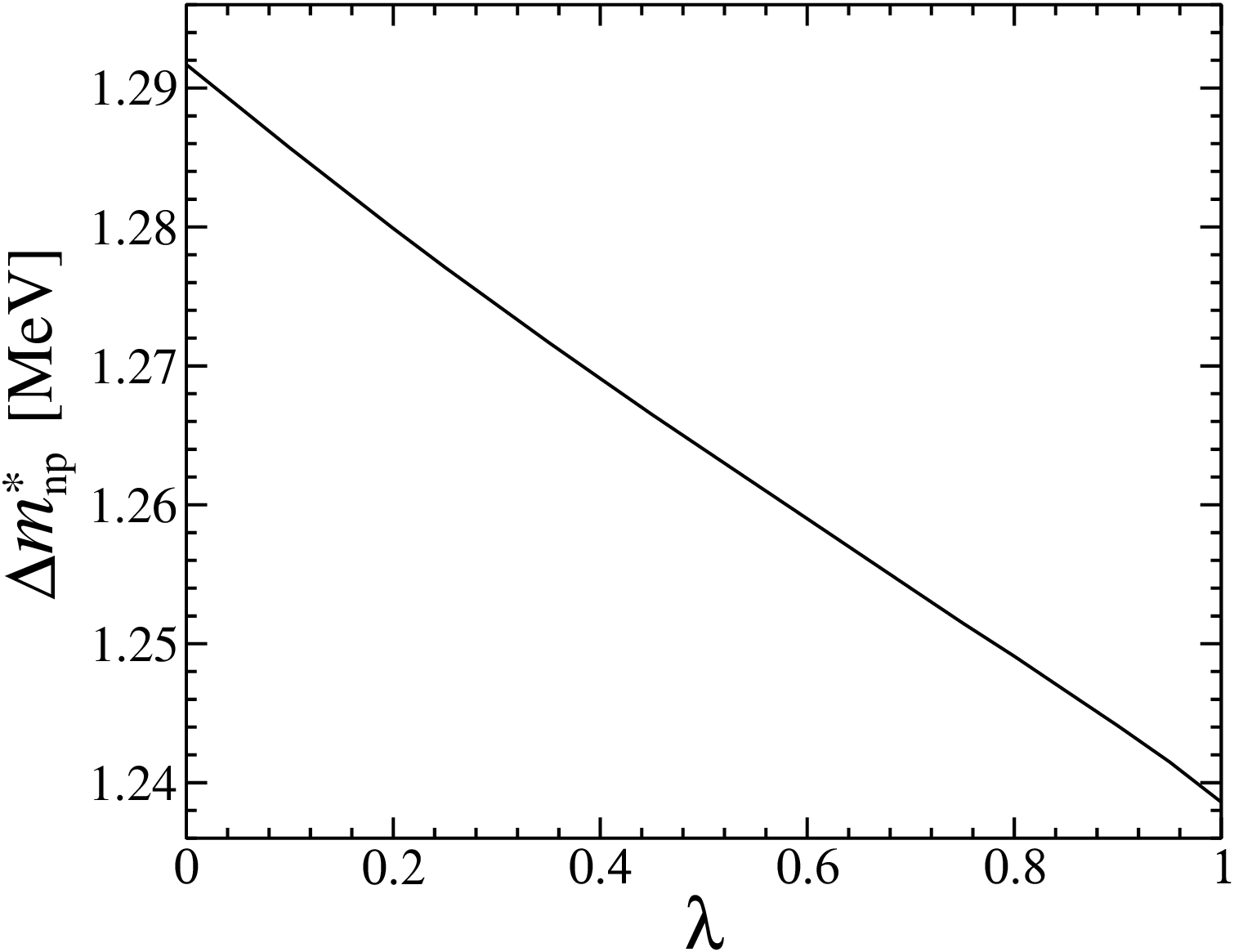}}}
\vspace{0.2cm}      
\caption{Density dependence of the 
total neutron-proton mass difference.
The abscissa represents $\lambda=\rho/\rho_0$,
while the ordinate represents  $\Delta m_{\rm np}^*$ in units of MeV.}
\label{fig3}
\end{figure}

\begin{figure}
\centerline{\resizebox{0.41\textwidth}{!}{\includegraphics{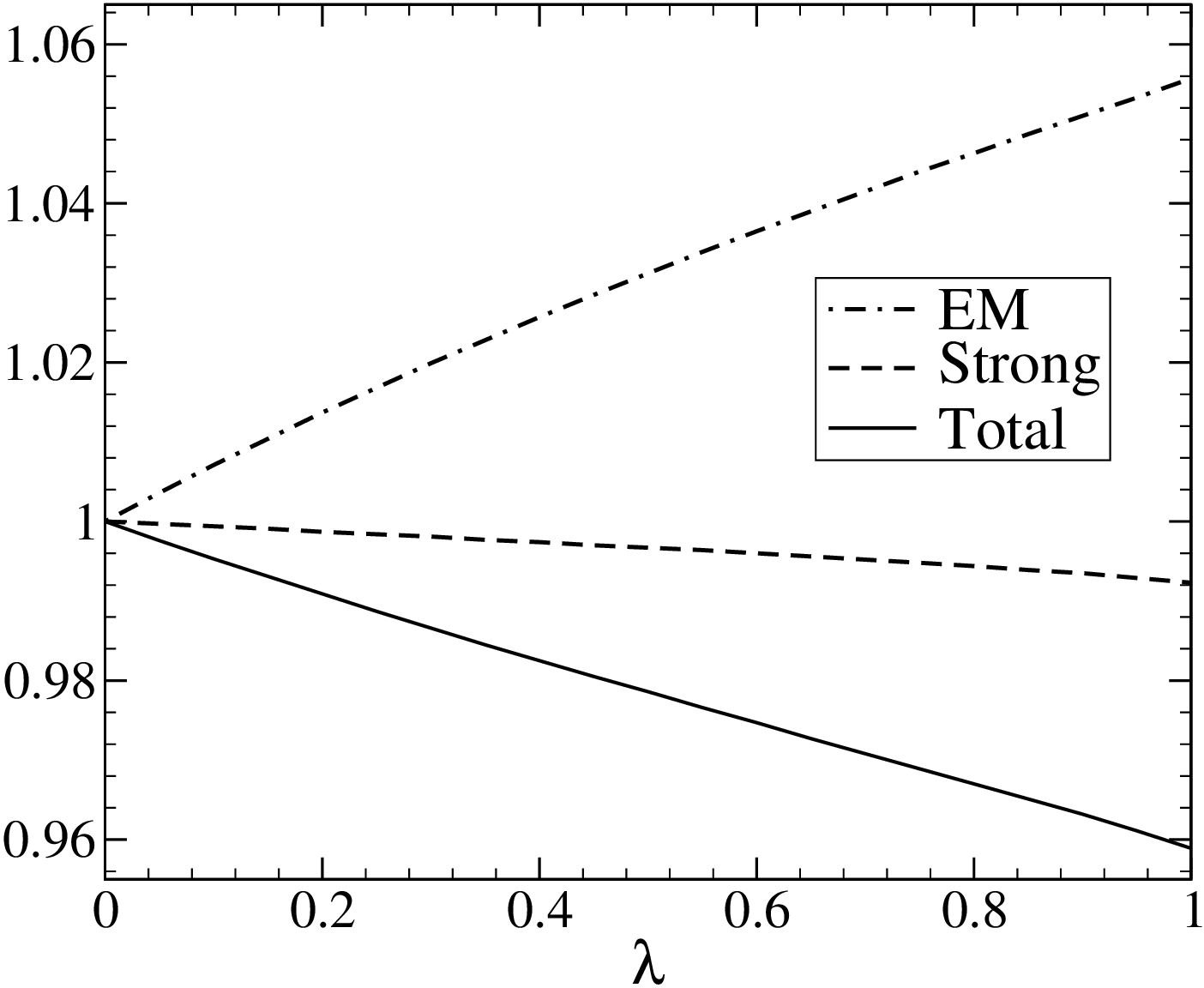}}}
\vspace{0.2cm}
\caption{Density dependence of the 
following neutron-proton mass differences,
normalized to their free space values:
${\Delta m_{\rm np}^{*(\rm EM)}}/{\Delta m_{\rm np}^{(\rm EM)}}$,
${\Delta m_{\rm np}^{*(\rm strong)}}/{\Delta m_{\rm np}^{(\rm strong)}}$,
${\Delta m_{\rm np}^*}/{\Delta m_{\rm np}}$ plotted by the dash-dotted,
dashed
and solid curves, respectively.
The abscissa represents $\lambda=\rho/\rho_0$,
while the ordinate represents the three dimensionless ratios.}
\label{fig4}
\end{figure}

In Fig.~\ref{fig3} the total neutron-proton mass
difference $\Delta m_{\rm np}^*$ in the nuclear medium is shown as function of
the nuclear density.
For  convenience, we present also, normalized to their free space values,
the neutron-proton mass differences
${\Delta m_{\rm np}^{*(\rm EM)}}/{\Delta m_{\rm np}^{(\rm EM)}}$,
${\Delta m_{\rm np}^{*(\rm strong)}}/{\Delta m_{\rm np}^{(\rm strong)}}$, and
${\Delta m_{\rm np}^*}/{\Delta m_{\rm np}}$,
in Fig.~\ref{fig4} by the dash-dotted, dashed and solid curves,
respectively.

{}From Fig.~\ref{fig4} one can see
that the change in the neutron-proton mass difference is only
$\sim$ 4\% at normal nuclear matter density and that the total value
$\Delta m_{\rm np}^*$ decreases in the nuclear medium as function of the
density. This result
is in qualitative agreement  with the result 
of the quark-meson  coupling model~\cite{Saito:1994tq}. At the
quantitative level, however, they differ: the 
absolute value of the quantity
$\Delta m_{\rm np}^*$, within the present approach, decreases about
0.05~MeV at normal nuclear matter density (see Fig.~\ref{fig3}),
while in Ref.~\cite{Saito:1994tq} the decrease is about 0.85~MeV.
One can conclude that in the Skyrme model, contrary to the result of the
work~\cite{Saito:1994tq}, $\Delta m_{\rm np}^*$ remains positive --  even at
high densities -- if  the 
surrounding environment  has an isospin-symmetric
structure. Note that in isospin-asymmetric matter this picture
may change.

The density dependence of the neutron-proton mass difference is approximately 
linear and can be parameterized as
$$
\Delta m_{\rm np}^* \simeq 1.3~\mbox{MeV}-C\,\rho/\rho_0\,.
$$
Our calculation shows that $C\simeq 0.05$~MeV, which is almost negligible
in comparison with the calculation~\cite{Schafer:1993xt}
 in the framework of QCD sum rules,
$C\simeq 1.1\div 1.7$~MeV.

\section{Summary and outlook}
\label{sec:summary}

We have investigated the isospin-breaking effects for
nucleons embedded into an isospin-symmetric nuclear
environment. In order to calculate these effects, a medium-modified
version of the Skyrme model, which also takes into account 
explicit isospin breaking,   has been proposed. Our
calculations within this framework show 
that,  with increasing density, 
the total neutron-proton mass difference as well as 
its strong and its electromagnetic part, separately, 
decrease only by a very small amount. 

A generalization of the present approach to  finite nuclei, \textit{i.e.} to 
calculations 
of the type  presented
in Ref.~\cite{Yakhshiev:2001ht}, would allow to evaluate the
isospin-breaking effects in mirror nuclei with the following qualifications:

\begin{itemize}
\item[{-}]
the strong part of the neutron-proton mass difference would separately
depend on the S- and P-wave medium functionals $\chi_{\rm s,p}^\pm$,
$\chi_{\rm s,p}^0$ and, consequently, would strongly depend on the
parameterization of the latter ones~\cite{Ericsonbook};
\item [{-}]
the medium functionals would explicitly depend on the neutron-proton density
distributions 
$\chi_{\rm s,p}^{\pm,0}=\chi_{\rm s,p}^{\pm,0}(\rho_{\rm n}+\rho_{\rm p},
\rho_{\rm n}-\rho_{\rm p})$~\cite{Ericsonbook},
such that the main property of mirror nuclei
$$ \mbox{}^A_Z {\mathbf{M}}_N \quad\Leftrightarrow\quad
 \mbox{}^A_{\tilde Z} \widetilde{\mathbf{M}}_{\tilde N}
=\mbox{}^A_{N} \widetilde{\mathbf{M}}_{Z}
$$
could be taken into account in a natural way;
\item [{-}]
additional effects  could arise due to an  $\omega$-de\-pendence of
the polarization operators
$\hat\Pi^{\pm,0}=\hat\Pi^{\pm,0}(\omega,\vec{k})$ in
Eq.~\re{Pol-op}\footnote{Note that in the present work, the polarization
operator, 
because of the form of Eq.~\re{lmb} and
the isospin symmetry of the surrounding environment, 
is simply given by
the static formula
$\hat\Pi^{\pm,0}= \hat\Pi(\omega=m_{\pi^0},\vec{k})$.}, \textit{e.g.} from the 
Weinberg-Tomozawa~\cite{Weinberg:1966kf,Tomozawa:1966jm}  
and the so-called range terms~\cite{Delorme:1992cn}, see also
Refs.~\cite{Thorsson:1995rj,Meissner:2001gz};
\item [{-}]
the evaluation of {\em local} isospin-breaking effects, 
according to the nucleon-position $\vec{R}$ as measured from the center 
of the nucleus~\cite{Yakhshiev:2001ht}, would lead to
additional possibilities.
\end{itemize}

The consequences of these additional modifications are the subject 
of future  studies.

\begin{acknowledgement}

We thank Norbert Kaiser for a useful correspondence. 
The work of U.T.Y.\ was supported by the
Alexander von Humboldt Foundation. A.M.R.\
thanks Prof.\,V.\,E.\,Kravtsov 
for hospitality during his stay at the ICTP, Italy,
the Deutscher Akademischer Austausch Dienst (DAAD) for a travel scholarship
to J\"ulich, Germany, and the Forschungszentrum
J\"ulich for hospitality.
Partial financial support from the EU Integrated Infrastructure
Initiative Hadron Physics Project (contract number RII3-CT-2004-506078),
by the DFG (TR 16, ``Subnuclear Structure of Matter'') and BMBF
(research grant 06BN411) is gratefully acknowledged.

\end{acknowledgement}

\end{document}